\documentclass[12pt,a4paper]{iopart}

\usepackage{iopams}  
\usepackage{amssymb}
\usepackage{epsfig,newlfont}
\usepackage{subfigure}
\usepackage[colorlinks, 
linkcolor=blue,
urlcolor=blue,
citecolor=blue,
bookmarksopen=false]{hype rref}
\usepackage[usenames,dvipsnames]{xcolor}
\usepackage{graphicx}
\usepackage{caption}

\newcommand{\bb}{\mathbf{b}}
\newcommand{\bB}{\mathbf{B}}
\newcommand{\bj}{\mathbf{j}}
\newcommand{\bu}{\mathbf{u}}
\newcommand{\epol}{\mathbf{e}_\theta}
\newcommand{\fsa}[1]{\langle{#1}\rangle}

\begin{document}

\title{{Damping of {radial electric field fluctuations} in the TJ-II stellarator}}

\author{J.~L. Velasco$^1$, J.~A. Alonso$^1$, I. Calvo$^1$,
  J. Ar\'evalo$^1$, E. S\'anchez$^1$, L. Eliseev$^2$, S. Perfilov$^2$,
  T. Estrada$^1$, A. L\'opez-Fraguas$^1$, C. Hidalgo$^1$ 
{and} the
  TJ-II team$^1$}

\address{$^1$ Laboratorio Nacional de Fusi\'on, {Asociaci\'on}
  EURATOM-CIEMAT, Madrid, Spain} \address{$^2$ RNC Kurchatov
  Institute, Moscow, Russia} \ead{joseluis.velasco@ciemat.es}

\begin{abstract}

  The drift kinetic equation is solved for low density TJ-II plasmas
  employing slowly varying, time-dependent profiles. This allows to
  simulate density ramp-up experiments and to describe from first
  principles the formation and physics of the radial electric field
  shear, {which is associated to the transition from electron to
    ion root}. We {show} that the range of frequencies of plasma
  potential fluctuations in which zonal flows are experimentally
  observed is {neoclassically undamped in a neighborhood of the
    transition}. This makes the electron root regime of
  {stellarators, close to the transition to ion root,} a
  propitious regime for the study of zonal-flow evolution. {We
    present simulations of {collisionless relaxation of zonal
      flows, in the sense of the Rosenbluth and Hinton test, that show
      an oscillatory behaviour in} qualitative agreement with the
    experiment close to the transition.}

\end{abstract}

\section{Introduction}\label{SEC_INTRO}

Sheared radial electric fields are generally accepted to play a
central role in confinement transitions in fusion
plasmas~\cite{burrell1997ExB,terry2000shear,wagner2007hmode}. In
non-quasisymmetric stellarators, the mean radial electric field is
expected to be determined by the ambipolar condition of neoclassical
fluxes~\cite{helander2008amb,calvo2013er}. Nevertheless,
turbulence-generated radial electric field fluctuations that display a
zonal character have been observed in several stellarators
\cite{fujisawa2004prl,pedrosa2008prl}. In this work, we discuss how
turbulent momentum fluxes can overcome the collisional (due to
neoclassical viscosity) and collisionless damping and modify the
$E\times B$ rotation in the vicinity of a neoclassical bifurcation of
the radial electric field~\cite{shaing1984stab}.

The TJ-II stellarator exhibits such a bifurcation in low density
plasmas with Electron Cyclotron Heating (ECH), see
reference~\cite{milligen2011transitions} and references therein. When a
critical density $n_{cr}$ is reached, a spontaneous confinement
transition takes place, associated with the formation of a shear
layer. Experimentally, one of its most salient features is the
emergence {of zonal-flow-like} electric potential
structures~\cite{pedrosa2008prl,alonso2012zf}. When the density decreases, the
reverse transition occurs at a similar $n_{cr}$, but the amplitude of
the {zonal flows} is smaller.

In this work, we solve the drift kinetic equation for slowly varying,
time-dependent profiles. The evolution of the mean radial electric
field is successfully described from first principles and we also
provide a fundamental explanation for a wealth of experimental
observations in the neighbourhood of the critical density. The key
quantity is the neoclassical viscosity~\cite{velasco2012prl}, which
goes smoothly to zero when the critical density is approached from
below. Since this viscosity acts as the restoring force of
{deviations of the radial electric field $E_r$ from}
ambipolarity, large $E_r$ excursions and, in particular, {zonal
  flow fluctuations can be {better} observed}. These predictions
are illustrated in this work by density ramp experiments that
exemplify the typical phenomena observed during the transition. The
completion of the picture requires to study the time evolution of
zonal flows in these plasmas. {This is done by simulating the
  collisionless linear damping of zonal flows with the gyrokinetic code
  EUTERPE~\cite{jost2001euterpe,kleiber2011euterpe}.} {These
  simulations are performed in a neoclassical background, and the
  predicted zonal flow oscillations are enhanced by the decrease of
  the radial electric field during the neoclassical bifurcation, in
  agreement with previous experimental observations.}

The rest of the paper is organized as follows. Section~\ref{SEC_EXP}
describes the experiments. Section~\ref{SEC_EQ} briefly introduces the
equations and our definition of neoclassical viscosity that we will
use to describe them; this is done in Section~\ref{SEC_NCLDT}.
Section~\ref{SEC_BLDT} then looks more into detail the backward
transition. Finally, {Section~\ref{SEC_GK} describes linear gyrokinetic
simulations in the plasma conditions of our experiment.} The
conclusions are drawn in Section~\ref{SEC_CONCL}.

\section{Experimental results}\label{SEC_EXP}

\begin{figure}
\begin{center}
  \includegraphics[angle=0]{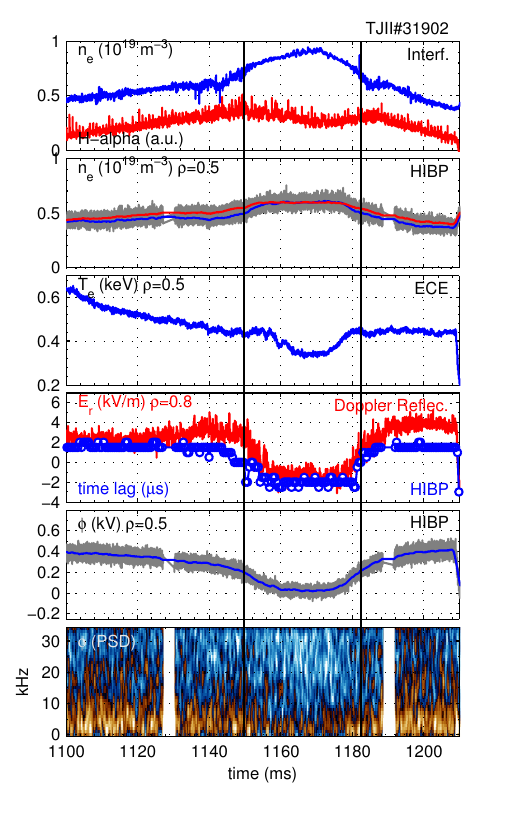}
\end{center}
\caption{From top to bottom, temporal evolution, for
  discharge \#31902, of the following quantities: $\overline{n_e}$ and
  $H_\alpha$ signal, $n_e(\rho=0.5)$, $T_e(\rho=0.5)$, $E_r(\rho=0.8)$
  measured with Doppler Reflectometry and time-lag in the HIBP signal
  (proxy for the sign of $E_r(\rho=0.5)$, see text), $\phi(\rho=0.5)$, and
  spectrogram of the potential fluctuations at $\rho=0.5$.}
\label{FIG_TRACES}
\end{figure}

In this section, we briefly describe experiments in which the main
phenomenology associated to the low-density transition of TJ-II is
present. These are the experimental results that we will try to
reproduce in Sections~\ref{SEC_NCLDT} and ~\ref{SEC_GK} {with}
simulations from first principles. Figure~\ref{FIG_TRACES} shows the
temporal traces of the relevant quantities in a representative
discharge. The line-averaged density $\overline{n}_e$ is provided by
interferometry. The local density and the plasma potential $\phi$ are
measured by Heavy Ion Beam Probe (HIBP) at several radial positions
($\rho\!=\!0.5$ in the selected discharge, with $\rho\!=\!r/a$ the
normalized radius, see below). The evolution of the local electron
temperature $T_e$ is taken from Electron Cyclotron Emission
(ECE). Both $T_e$ and $n_e$ are calibrated with Thomson Scattering
(TS). The radial electric field $E_r$ at $\rho\!\approx\!0.8$ is taken
from Doppler Reflectometry (DR).

The plasma density is increased at a constant rate by means of gas
puffing and, for a constant ECH power input, $T_e$ decreases. When
$\overline{n}_e$ reaches a value of about $0.6 -0.7\times
10^{19}\,$m$^{-3}$ an spontaneous confinement transition takes
place. The particle confinement time increases~\cite{tabares2001taup},
as indicated by the $H_\alpha/n_e$ ratio, which drops (note also the
change of slope in the $\overline{n}_e(t)$ curve). At the same time,
$E_r$ changes from positive to negative. The local reversal of $E_r$
is observed by the DR; the reversal of $E_r$ at $\rho\!>\!0.5$ causes
the drop in the plasma potential $\phi(\rho\!=\!0.5)$ measured by
HIBP.  Since $\phi(\rho\!=\!1)$ is set to 0$\,$V, $\phi(\rho=0.5)$
contains information about $E_r(0.5\!<\!\rho\!<\!1.0)$, which makes it
difficult to use this measurement to detect the local reversal of
$E_r$. Instead, we use the time-lag in the detection of density
fluctuations (which we assume advected by the $E\times B$ {flow}) in two
poloidally-separated sample volumes of the HIBP, see
figure \ref{FIG_TRACES}. For discharges in which the position of
measurement agree, we have checked that this procedure gives us
information on the sign of $E_r$ in good agreement with DR.

We will see in the next section that the reversal of $E_r$ typically
starts around the region where the density gradient is maximum, and a
sheared $E_r$ appears which then propagates across the entire plasma
radius at a speed of the order of several m/s~\cite{happel2008twostep}
given by the density ramp. Then, further increase in $\overline{n}_e$
does not modify qualitatively $E_r$. In the selected discharge, the
density ramp is relatively fast and this dynamics cannot be resolved:
we see that $E_r$ changes sign almost simultaneously in all the region
$\rho\!>\!0.5$.

\begin{figure} 
\begin{center} 
\includegraphics[angle=0]{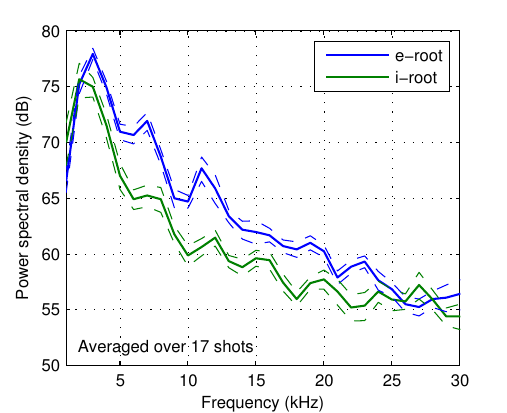}
\includegraphics[angle=0,width=\columnwidth]{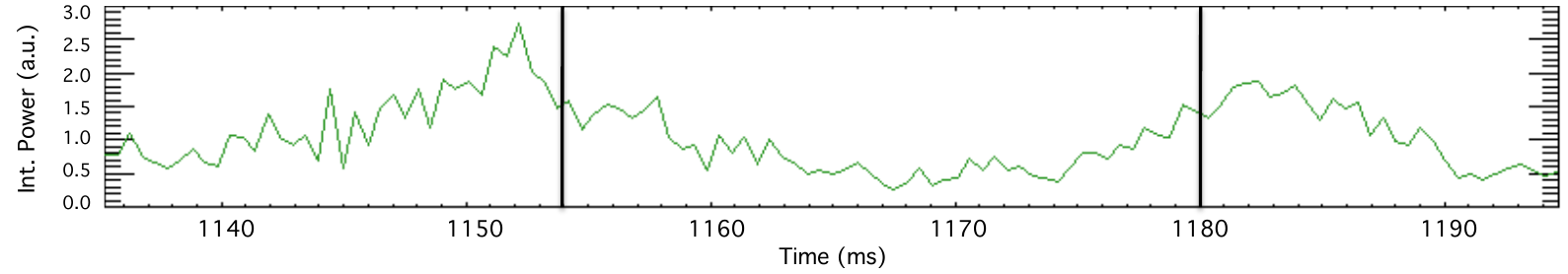}
\end{center} 
\caption{(top) Average spectrum of the potential fluctuations measured
  at $\rho\!=\!0.5$ by means of HIBP for $n\!\lesssim\!n_{cr}$
  (electron root, blue) and $n\!>\!n_{cr}$ (ion root, green); (bottom)
  integrated power spectra measured by Doppler reflectometry at
  $\rho\!=\!0.8$ for the discharge of figure \ref{FIG_TRACES}.}
\label{FIG_SPECTRA} 
\end{figure}

Figure~\ref{FIG_TRACES} also shows the evolution of the spectrum of
low-frequency (tens of kHz) potential fluctuations measured by HIBP as
the density ramp takes place. The amplitude of the fluctuations is
smaller when $E_r$ is negative, and a maximum of fluctuations is
detected for densities slightly lower than $n_{cr}$, although it is
difficult to spot in figure \ref{FIG_TRACES}. To show that these
features are systematic, we plot in figure \ref{FIG_SPECTRA} (top) the
average spectrum of the potential fluctuations at $\rho\!=\!0.5$ for
two time windows, one corresponding to $n\!\lesssim\!n_{cr}$ and the
other to $n\!>\!n_{cr}$, averaged over 17 reproducible discharges. The
peak in fluctuations close below $n_{cr}$ is {best} seen in
figure \ref{FIG_SPECTRA} (bottom), where we show the level of $E_r$
fluctuations ($\omega\!<200\,$kHz, with a large contribution of
$\omega\!<50\,$kHz frequencies) measured by DR. for the same discharge
of figure \ref{FIG_TRACES}.

It is worth noting that in
Refs.~\cite{hidalgo2004pre,pedrosa2007sheared} this peak of
low-frequency potential fluctuations has been associated to {Long
  Range Correlated (LRCed)} electrostatic potential structures that
grow when approaching the
critical density and are a proxy to zonal flows.\\

Figures~\ref{FIG_TRACES} and figure \ref{FIG_SPECTRA} also include the
evolution of the plasma when the gas puffing rate is reduced and the
density is ramped down. The particle confinement time goes back to
pre-transition levels, $E_r$ goes back to positive and the potential
fluctuations grow again. A peak in the level of fluctuations, of
smaller amplitude, is observed {\em after} the transition, see
figure \ref{FIG_SPECTRA} {(bottom), consistently} with previous
observations~\cite{milligen2011transitions}. No hysteresis {in
  $E_r$} is evident from Figure~\ref{FIG_TRACES}, but a more detailed
study will be made in Section~\ref{SEC_BLDT}.

\section{{Neoclassical equations and simulations}}\label{SEC_EQ}

Most of the phenomenology described in the previous section will
appear in this work as a natural consequence of a neoclassical
bifurcation at $n_{cr}$. For this reason, we will briefly derive the
equation for the evolution of the {radial electric field} that
we will solve in the next sections.  We start from the momentum
balance equation~\cite{helander2008amb} summed over {species,}
\begin{equation}
  m_i\frac{\partial (n\bu)}{\partial t} + \nabla\cdot\Pi_i +
  \nabla\cdot\Pi_e = \bj\times\bB\,.~\label{EQ_MOMBAL}
\end{equation}
Here, $\bu$ is the ion flow tangent to flux surfaces, $\Pi_s$ is the
viscosity tensor, $\Pi_s =
m_s\int\mathbf{v}\mathbf{v}f_s(\mathbf{x},\mathbf{v},t)
d^3\mathbf{v}$, or momentum flux of species $s$, $f_s$ its
distribution function, and $\bj\times\bB$ is the Lorentz force. The
projection of the latter over the flux-surface (proportional to the
radial current), is set to zero unless otherwise stated. This is
implied by quasineutrality ($\nabla\cdot\bj = 0$), but a net radial
plasma current can be induced e.g. in plasma biasing experiments. We
model a quasineutral plasma consisting of singly charged ions and
electrons ($n_e\!=\!n_i\!=\!n$). Note that we have dropped the inertia
of the electrons, given their much lower mass, $m_e/m_i\!\ll\!1$, but
kept the electron viscosity tensor as it cannot be neglected in our
low-$n$, high-$T_e$ ECH-heated plasmas~\cite{velasco2012er}. We work
in Hamada magnetic coordinates $(\psi, \theta, \xi)$, and follow the
notation of reference~\cite{alonso2012zf}. The lowest order incompressible
ion flow is conveniently written {as}

\begin{equation}\label{EQ_FLOW} \bu =
  2\pi\left(\frac{p_i'(\psi)}{ne} + \phi'(\psi)\right)\epol +
  \Lambda(\psi)\bB\,.  
\end{equation} 

The flux surface label $\psi$ is the toroidal magnetic flux, $p_i$ is
the ion pressure, $e$ is the elementary charge and the prime stands
for {differentiation with respect to} {the argument}. The poloidal
basis vector $\epol\!=\partial\mathbf{r}/\partial\theta$
satisfies $\epol\times\bB = (2\pi)^{-1}\nabla\psi$, $\nabla\cdot\epol
= 0$ and, for a currentless stellarator, $\fsa{\epol\cdot\bB} = 0$.
{We note that the form of the flow in {equation} (\ref{EQ_FLOW}) is
  derived under a transport ordering that reduces particle
  conservation to the incompressibility condition $\nabla\cdot\bu=0$.}
The first term on the {right-hand side} of equation (\ref{EQ_FLOW})
contains the diamagnetic and $E\times B$ perpendicular {flows,
  together with the parallel Pfirsch-Schl\"uter flow}. The term
$\Lambda\bB$ is the ion bootstrap flow~\cite{velasco2011bootstrap}. If
we project equation (\ref{EQ_MOMBAL}) along $\epol$ and take {
  flux-surface-average, denoted by $\fsa{\cdot}$,} we obtain our
evolution equation for the radial electric {field,}
\begin{eqnarray}
  \frac{\partial E_r}{\partial t} &=&
  \frac{1}{n}\frac{\partial}{\partial t}\left(\frac{p_i'(r)}{e}\right)
  - E_r\frac{1}{n}\frac{\partial n}{\partial t} +
  \nonumber\\&+&\frac{(\psi'(r))^2}{4\pi^2
    {m_i}n\fsa{\epol\cdot\epol}}\left(e(\Gamma_e-\Gamma_i)+\fsa{\bj\cdot\nabla
      r}\right){.}\label{EQ_ER} 
\end{eqnarray} 
{Here,} $E_r\equiv -\phi'(r)$ and the radius $r$ is a geometric
flux label defined in terms of the volume $V(r)\!\equiv\!\pi
r^2{L_{\textrm{\scriptsize ax}}}$, where
${L_{\textrm{\scriptsize ax}}}$ is the length of the magnetic
axis and {$V(a)$ is the total plasma volume,} with $a$ the minor
radius{. We} have obtained the radial particle fluxes from
$\Gamma_s\!=\!-\frac{2\pi}{q_s\psi'(r)}\fsa{\epol\cdot\nabla\cdot\Pi_s}${~\cite{hirshman1981ff}.}
The viscosity tensor can be split into a neoclassical part, given by
the gyrotropic pressure tensor, and an anomalous {contribution,}

\begin{equation} \Pi_s = \Pi_s^{NC} + \Pi_s^{an} = p_{s\|}\bb\bb +
  p_{s\perp}\left(\mathbf{I}-\bb\bb \right) +
  \Pi_s^{an}\,.\label{EQ_PI} 
\end{equation} 
As mentioned above, in non-quasisymmetric confining magnetic
topologies, the leading order contribution to equation (\ref{EQ_ER}) is
$\fsa{\epol\cdot\nabla\cdot\Pi_s^{NC}}$, being much larger than
$\fsa{\epol\cdot\nabla\cdot\Pi_s^{an}}$, which will be therefore
neglected~\cite{helander2008amb,calvo2013er}.

\begin{figure} 
  \begin{center} 
    \includegraphics[angle=0,width=0.5\columnwidth]{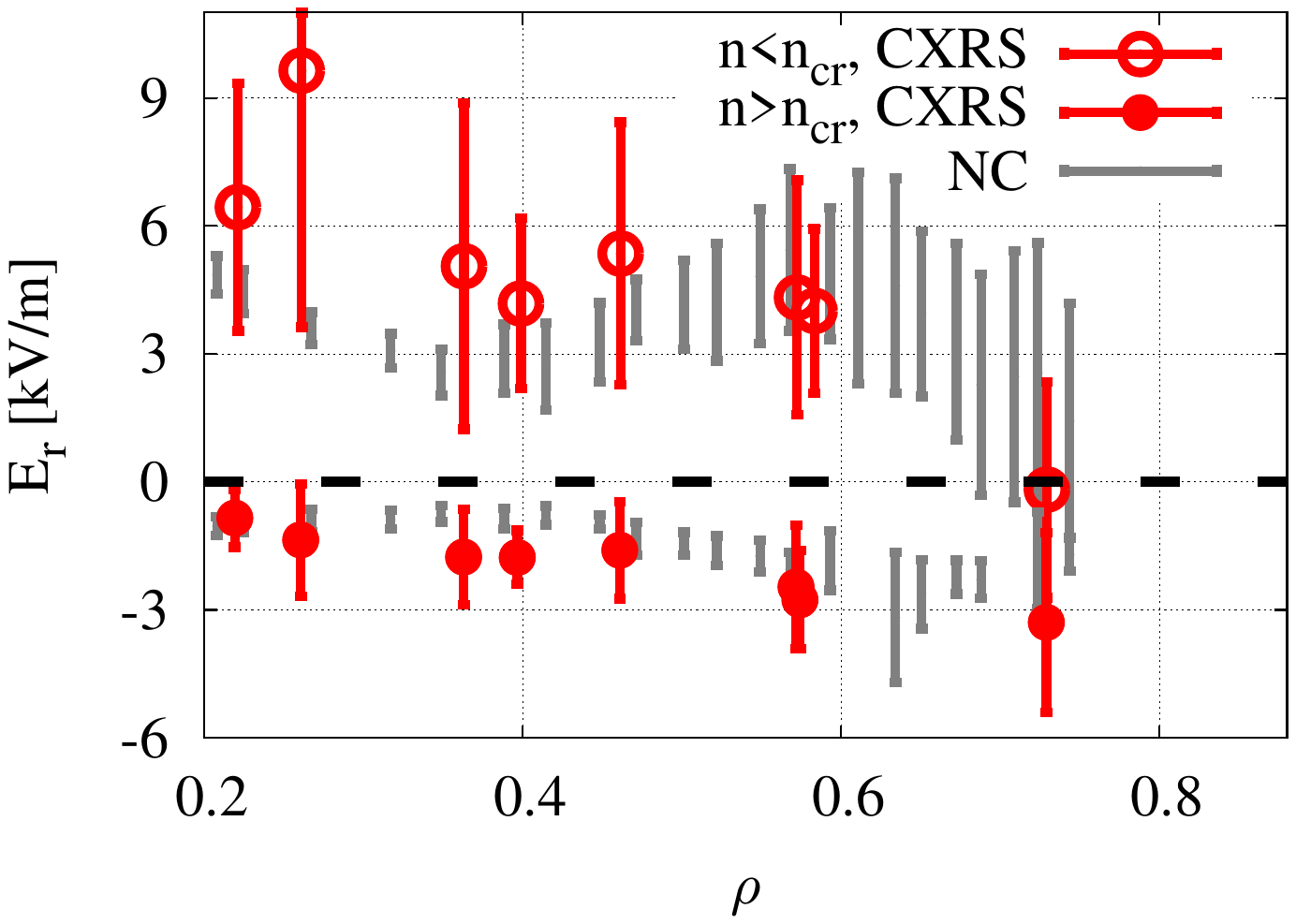} 
    \includegraphics[angle=0,width=0.5\columnwidth]{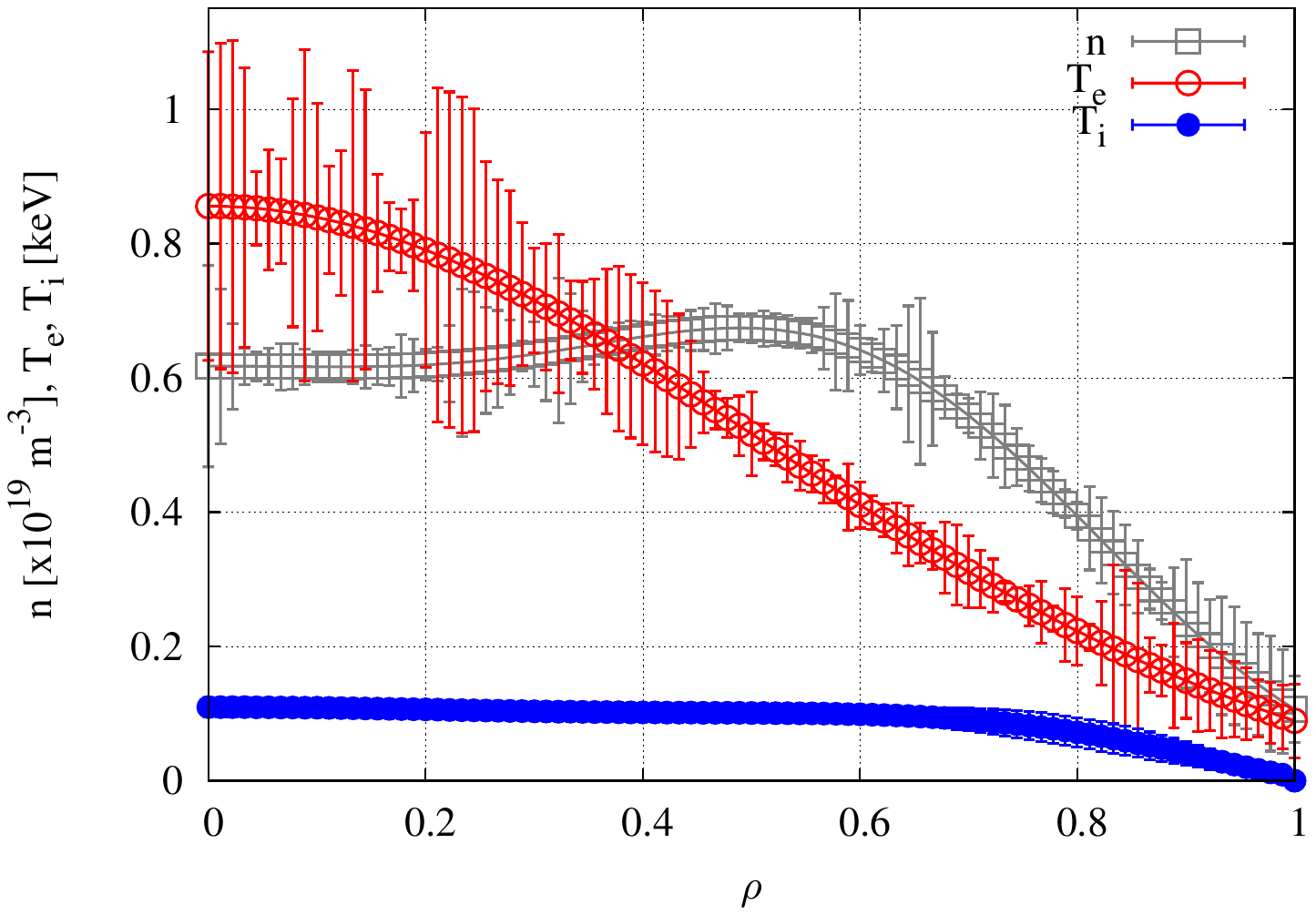} 
    \includegraphics[angle=0,width=0.5\columnwidth]{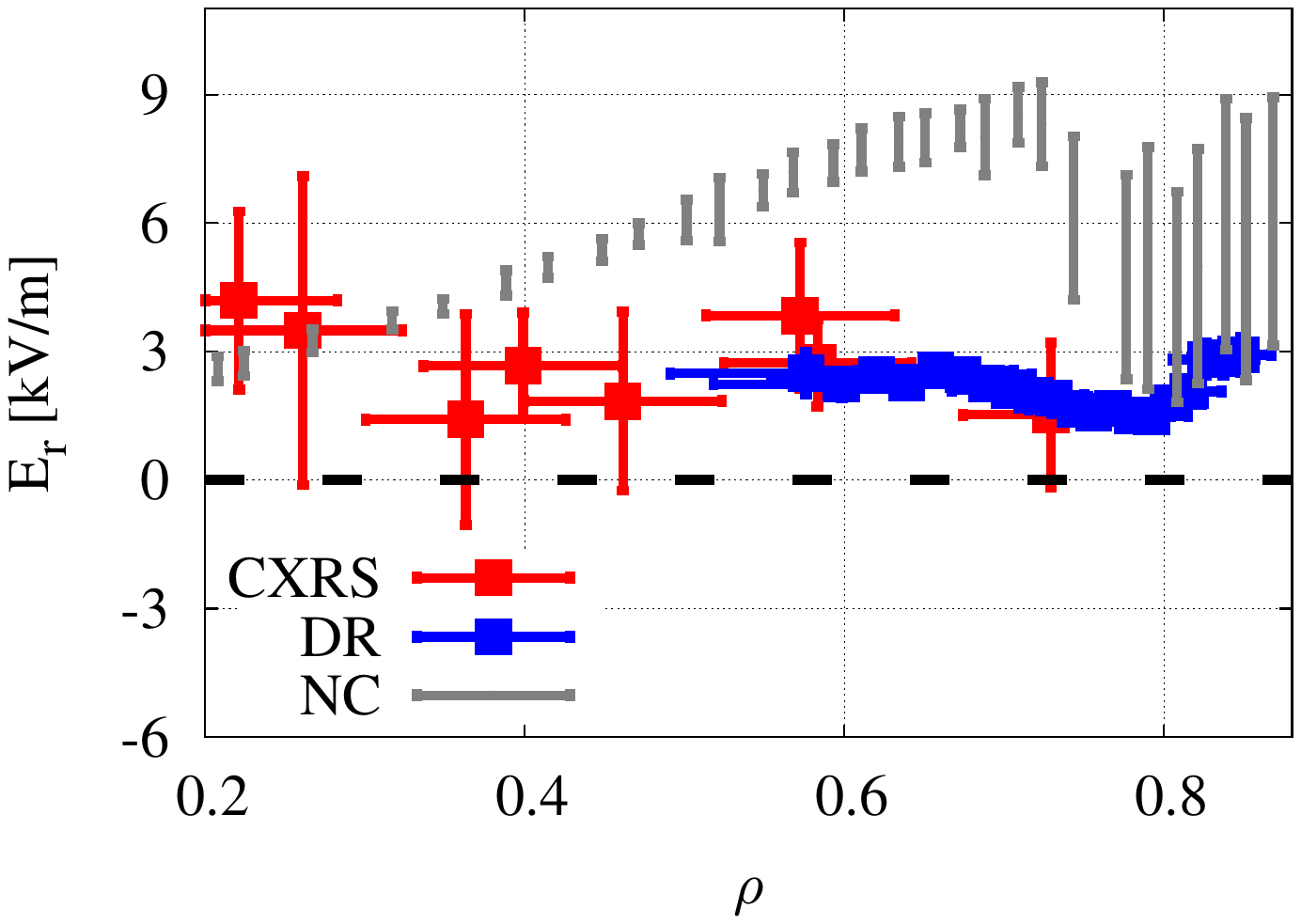} 
  \end{center} 
  \caption{(top) Radial electric field profile measured
    by CXRS and compared to neoclassical calculations for
    $n\!<\!n_{cr}$ and $n\!>\!n_{cr}$; (center) plasma profiles of
    discharge \#32599 , $n\!\lesssim\!n_{cr}$; (bottom) radial electric
    field profile measured by CXRS and DR compared to neoclassical
    calculations for a set of similar discharges that includes
    \#32599.}
  \label{FIG_CXRS} 
\end{figure}

We will use the Drift Kinetic Equation Solver
(DKES)~\cite{hirshman1986dkes} complemented with momentum-conserving
techniques~\cite{maassberg2009momentum} to evaluate the pressure
anisotropy in a realistic TJ-II magnetic field equilibrium
configuration in the parameter range in the vicinity of the
transition. Details of the numerical calculation can be found in
reference~\cite{velasco2012er}. In order to check the accuracy of our
numerical results, we first compare the predictions to {Charge eXchange
Recombination Spectroscopy (CXRS)} measurements and Doppler Reflectometry of
steady state $E_r$ in real discharges. We start from equation
\ref{EQ_ER} and set to zero the time-dependent terms and the Lorentz
force, i.e. we solve $E_r$ from the ambipolar equation for the radial
neoclassical fluxes, $[\Gamma_e-\Gamma_i](E_r)\!=0\,$.

Plasmas
well below (above) $n_{cr}$ with $E_r$ positive (negative) in all
plasma radius have been thoroughly studied in
reference~\cite{arevalo2013flows1} and the predictions show good
qualitative and quantitative agreement with CXRS measurements. For
reference, we show some of the results in figure \ref{FIG_CXRS} (top). A
similar exercise was performed already in
reference~\cite{zurro2006rotation}, in which passive spectrometry
measurements were in qualitative agreement with neoclassical
calculations. 

Nevertheless, as we will see later when studying the ambipolar
equation of these plasmas, the closer one gets to $n_{cr}$, the more
challenging it is to calculate the precise value of $E_r$. This can
already be seen in figure \ref{FIG_TRACES} where, close to $n_{cr}$,
$E_r$ varies much faster than $n_e$ or $T_e$. This will set a limit to
the accuracy of our quantitative predictions. Both measurements and
predictions include results from a set of 5 reproducible discharges,
one of which is shown in figure \ref{FIG_CXRS} (center), and the results
of our comparison are presented in figure \ref{FIG_CXRS}
(bottom). According to the measurements, $E_r$ is positive in all the
plasma radius. At $\rho\!\approx\!0.8$, the DR measures a minimum of
$E_r$, with a double-shear layer at both sides of it. It is precisely
at this radial position where the reversal of $E_r$ typically
starts~\cite{happel2008twostep} which means that indeed the plasma is
very close to the transition.

The predicted neoclassical $E_r$ is positive, in agreement with the
experiment. We observe the double-shear-layer forming although at a
slightly outer radial position. Finally, our calculations overestimate
$E_r$ in the region $0.5\!<\rho\!<\!0.8$. The results are consistent
with an underestimation of the ion radial flux: {our calculation
  is radially local;} if finite-orbit-width effects were included in
the calculation of the ion radial flux~\cite{satake2006fortec3d}, the
theoretical predictions would probably come closer to the
experiment. This is currently under investigation.

It is now clear that, although we probably will overestimate $n_{cr}$,
our ordering assumption is valid for capturing {the behaviour of
  $E_r$} during the low density transition: the leading term in the
radial current is neoclassical and sets the {value of $E_r$};
turbulent contributions will be present and will produce excursions of
$E_r$ from its ambipolar value, see figure \ref{FIG_TRACES} (bottom)
and~\ref{FIG_SPECTRA}. {Let us} {now} {show that these fluctuations can
  also, to some extent, be discussed in terms of equation
  (\ref{EQ_ER}). First,} {we} {expand the radial neoclassical current around
  the ambipolar radial electric field $E_r^0$,
\begin{equation}
  [\Gamma_e-\Gamma_i](E_r)\!=-\mu_p(E_r-E_r^0)+O((E_r-E_r^0)^2).
\label{EQ_VISCO} 
\end{equation}
The neoclassical viscosity, $\mu_p$, is defined as the linear
coefficient in this expansion. If we neglect the slow variations in
$n$ and $T_e$, equation (\ref{EQ_ER}) yields
\begin{eqnarray} \frac{\partial \delta E_r}{\partial t} &&\approx
  \frac{e(\psi'(r))^2}{4\pi^2 mn\fsa{\epol\cdot\epol}}
  \Bigg[\mu_p\delta E_r -\frac{\fsa{\bj\cdot\nabla r}}{e}\Bigg]
  \nonumber\\&&=-\nu_p\delta E_r
  +\check{j_r},\label{EQ_NUP} 
\end{eqnarray}
where $\delta E_r = E_r - E_r^0$, and constants have been absorbed} {in
$\check{j_r}$ and in the viscosity $\nu_p$, which now has dimensions of frequency}. {Equation~\ref{EQ_NUP} is then meant to model the
evolution of dynamically incompressible $E\times B$ flow fluctuations,
$\tilde{\bu} = -(2\pi/\psi'(r))\delta E_r\epol$, with the perpendicular flow
$\tilde{\bu}_\perp = \delta E_r\nabla r\times \bB/B^2 = -(2\pi/\psi'(r))\delta
E_r{\epol}_\perp$ instantly accompanied by a parallel
Pfirsch-Schl\"uter flow $\tilde{\bu}_\| = -(2\pi/\psi'(r))\delta
E_r{\epol}_\|$. This results in the inertia factor
$\fsa{\epol\cdot\epol}\approx r^2(1+2/\iota^2)$, with the $2/\iota^2$
term coming from the parallel flow
response~\cite{Hallatschek2007}. Typical rotational transform values
in the edge of TJ-II are $\iota(a)\sim 1.5$ which do not substantially
modify $\nu_p$ values.  The assumption of dynamically incompressible
$E\times B$ flow fluctuations is justified for fluctuation frequencies
below the acoustic characteristic frequency of GAM oscillations. As shown
in Section~\ref{SEC_GK}, this frequency is about $40$ kHz for the
TJ-II plasmas under study, to be compared with typical zonal flow
frequencies $\lesssim 10$kHz \cite{pedrosa2007sheared, alonso2012zf}.}

In reference~\cite{velasco2012prl} $\check{j_r}$ represented the radial
current driven by an electrode, and the behaviour of $E_r$ during
biasing experiments~\cite{carralero2012suscep,pedrosa2007sheared} was
successfully reproduced. In this work, we will use it to understand
the behaviour of $E_r$ fluctuations during the density scan:
$\check{j_r}$ will be the driving term of fluctuations (e.g. Reynolds
Stress) and $\nu_p$ a restoring force towards neoclassical
ambipolarity. To make the argument more precise, we Fourier transform
equation (\ref{EQ_NUP}) and multiply times the complex-conjugate for
time scales faster than that of the density ramp, i.e.,
$\omega\!>\!\partial_t\log(E_r^0)\!\sim\!\partial_t\log(n)\!\sim\!10\,$Hz,
and {obtain
\begin{equation}
  |\hat{\delta E_r}(\omega)|^2 = \frac{1}{\nu_p^2+\omega^2}|\hat{j}(\omega)|^2\,.\label{EQ_OMEGA}
\end{equation}
Equation} \ref{EQ_OMEGA} shows that the amplitude of the fluctuations
{$\hat{\delta E_r}(\omega)$} driven by a given broadband
turbulent forcing $\hat{j}(\omega)$ is modulated by the neoclassical
viscosity, which damps fluctuations of frequencies lower than
$\nu_p$. We will see in the next section that, for our plasma
conditions, {$\nu_p\!\sim\!1 - 10 $ kHz}, of the order of the
relevant frequencies in figure \ref{FIG_SPECTRA}.

\begin{figure} 
\begin{center} 
\includegraphics[angle=0,width=1.0\columnwidth]{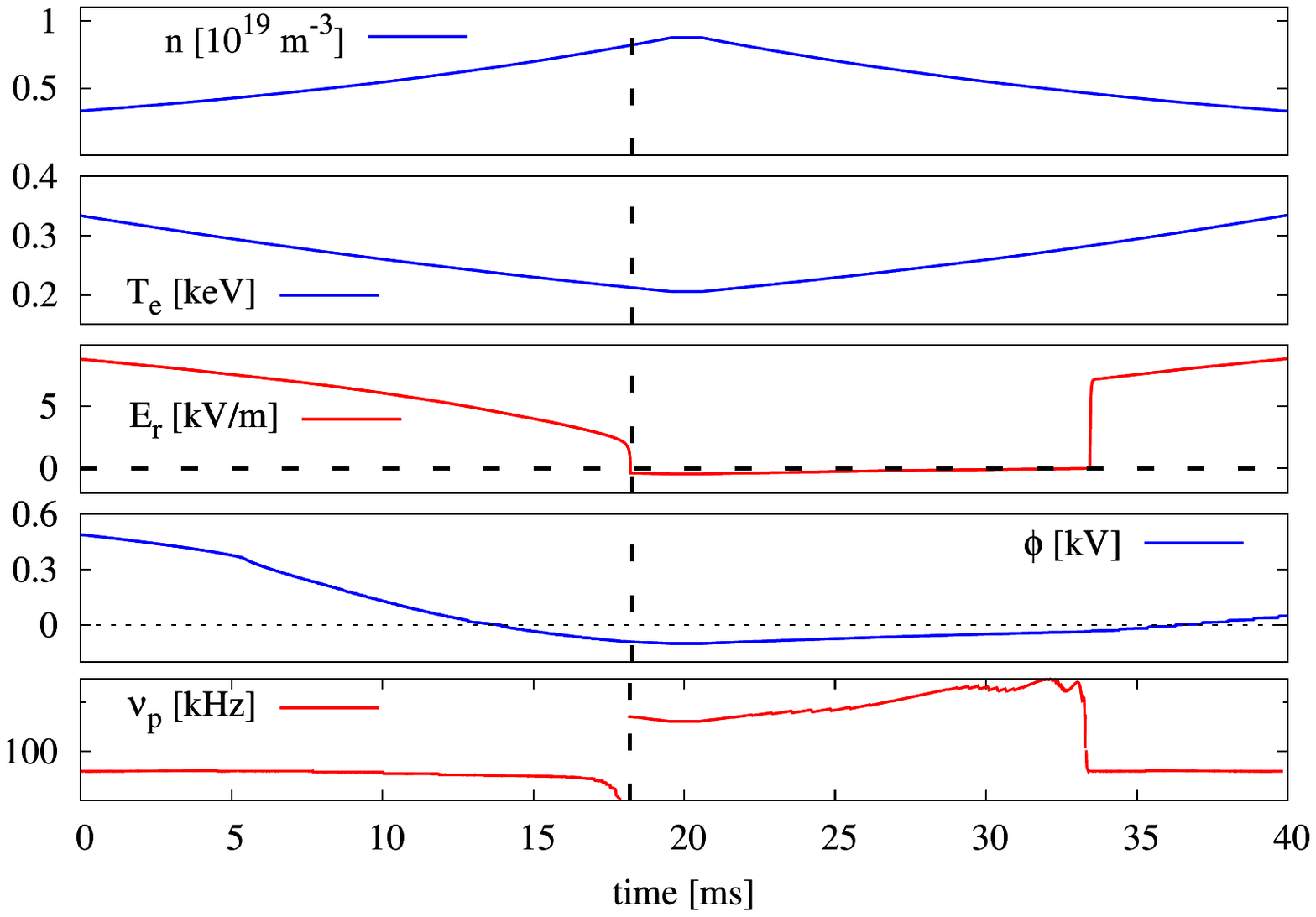} 
\end{center} 
\caption{From top to bottom, temporal evolution during
  the numerical ramp of the following quantities at $\rho\!=\!0.7$ :
  $n_e$, $T_e$, $E_r$, $\phi$ {, and $\nu_p$}.}
\label{FIG_EVOL}
\end{figure}

\begin{figure} 
  \begin{center} 
    \includegraphics[angle=0]{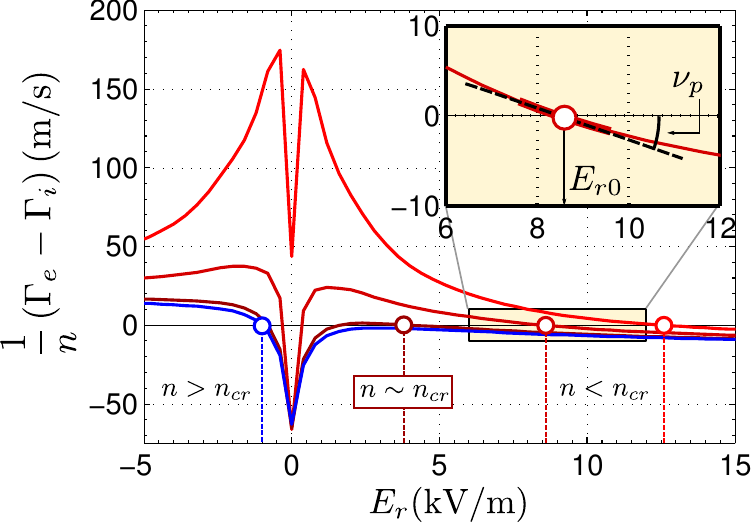} 
  \end{center} 
  \caption{Ambipolar equation at $\rho\!=\!0.7$ for
    several representative times; (inset) sketch of the estimation of
    the viscosity.}
  \label{FIG_AMB} 
\end{figure}

\section{Neoclassical description of the low-density
  transition}\label{SEC_NCLDT} 

We now perform a dynamical neoclassical calculation of the formation
of the sheared $E_r$. We therefore {use} equation (\ref{EQ_ER}) and {try to} simulate the density ramp-up and ramp-down of
{figure \ref{FIG_TRACES}}. We take smoothed plasma profiles similar to those
of figure \ref{FIG_CXRS} (top) and make them evolve {according to}
\begin{equation}
  \frac{1}{n}\frac{\mathrm{d}n}{\mathrm{d}t}={\pm 50\,\mathrm{s}^{-1}}\,,\quad \frac{1}{T_e}\frac{\mathrm{d}T_e}{\mathrm{d}t}={\mp 25 \, \mathrm{s}^{-1}}\,,\quad\frac{\mathrm{d}T_i}{\mathrm{d}t}=0\,,
\end{equation}
where the signs depend on the stage of the ramp. Note that a
simulation of the evolution of $n(\rho,t)$, $T_e(\rho,t)$ and
$T_i(\rho,t)$ is not required for the description of the formation of
the $E_r$-shear of TJ-II; the existence of anomalous fluxes and
sources is implicitly accounted for in the choice of $n(\rho,t)$,
$T_e(\rho,t)$ and $T_i(\rho,t)$.

{For low} (high) density, the predicted neoclassical $E_r$ is
positive (negative) in all the plasma radius, as in
{figure \ref{FIG_CXRS}} and reference~\cite{arevalo2013flows1}. The
evolution of the $E_r$ profile between these two situations (i.e.,
formation and evolution of the $E_r$-shear) was discussed in
reference~\cite{velasco2012prl}. Since our calculation is radially local,
we focus here in one representative radial position, $\rho\!=\!0.7$.

The result of the numerical density scan at this position is presented
in figure \ref{FIG_EVOL}, which is to be compared to
figure \ref{FIG_TRACES}. As $n$ rises and $T_e$ decreases, $E_r$ becomes
less positive. Then, for $n_e\!\approx\!0.7\times
10^{19}\,$m$^{-3}\,$ and $T_e\!\approx\!200\,$eV, there is a change
of root: $E_r$ goes from positive to negative in a time-scale of
several tens of $\mu$s. Further increase in $n$ then causes $E_r$ to
be more negative. The plasma potential $\phi$ evolves according to
$E_r$, and {therefore decreases during the transition.}

In figure \ref{FIG_EVOL} (bottom) we also plot the time evolution of
the neoclassical viscosity $\nu_p$. When the plasma is in the ion
root, the viscosity is much larger and, according to equation
(\ref{EQ_OMEGA}), the fluctuations have much smaller amplitude, as in
figure \ref{FIG_TRACES}. Immediately before the transition, a peak of
the low-frequency $E_r$ fluctuations is predicted because of the
vanishing $\nu_p$, which we discuss below.

We observe exactly the opposite behaviour ($E_r$ becomes less negative
and more positive, $\phi$ evolves accordingly, and the level of
fluctuations grows again) during the density ramp-down. The only
difference is the predicted values of $n_{cr}$ for the backward
transition, which differs from that of the forward transition. No
vanishing of $\nu_p$ is predicted. These issues will be further
discussed in Section~\ref{SEC_BLDT}.

Since the characteristic time scale of evolution of $E_r$
{(10$^{-5}$ s)} is much faster than that of the plasma profiles
{(10$^{-2}$ s)}, it is meaningful to discuss the results of
figure \ref{FIG_EVOL} in the light of the ambipolar equation of the
neoclassical radial fluxes for several selected plasmas of our density
scan. In figure \ref{FIG_AMB}, we show the radial neoclassical current
as a function of $E_r$ at $\rho\!=\!0.7$ for several relevant instants
of the density ramp.  This figure is characteristic of the
low-collisionality regime of stellarators~\cite{shaing1984stab}. For
very low collisionalities ($n\!\ll\!n_{cr}$) the only solution is a
positive $E_r$ (electron root); for higher collisionalities
($n\!\gg\!n_{cr}$) the only solution is a negative $E_r$ (ion
root). In these two situations, an increase in collisionality leads to
less positive or more negative $E_r$. In the electron root regime, the
variation of $E_r$ is larger, due to the smaller slope of the curve
$[\Gamma_e-\Gamma_i](E_r)$. For intermediate collisionalities, three
solutions of the ambipolar equation, one of them unstable, coexist. We
will define {the critical density $n_{cr}$ (respectively, the
  critical temperature $T_{cr}$) as the local density (respectively,
  the local electron temperature) at} which one of the two stable roots
disappears, so that $E_r$ {`jumps'} to the other solution. We discuss
about this criterion in Section~\ref{SEC_BLDT}.

The qualitative behaviour of the viscosity can be extracted from the
slope of the curves in figure \ref{FIG_AMB} (inset). It is larger in
the ion root regime than in the electron root, thus allowing for
smaller fluctuations {$\delta E_r$} provided that the forcing
$\check{j_r}$ does not change. According to figure \ref{FIG_EVOL}
(bottom) $\nu_p$ takes values of a few tens of kHz in the electron
root regime and {around} 100$\,$kHz in ion root.  Following
equation (\ref{EQ_OMEGA}), fluctuations of frequencies of about
10$\,$kHz will be more damped in the ion root. For higher-frequency
harmonics, equation (\ref{EQ_OMEGA}) becomes {$|\hat{\delta
    E_r}|/|\hat{j_r}|\!\sim\!\omega^{-1}$} and no differences are
expected between the two regimes. These predictions are in qualitative
agreement with figure \ref{FIG_SPECTRA} (top). {Figure
  \ref{FIG_SPECTRA} (bottom) shows that at the density for which the
  electron root disappears the viscosity vanishes, allowing for even
  larger {fluctuations $\delta E_r$}. Note that as the transition
  approaches and $\nu_p$ decreases, higher-order terms in the
  viscosity tensor of equation (\ref{EQ_PI}) (including those that
  depend on the $E_r$ shear) start to count in the evolution of
  $E_r$. Anyway, the experimental results of figures \ref{FIG_TRACES}
  and \ref{FIG_SPECTRA} prove that the reduction of $\nu_p$, that we
  predict without including such higher-order terms, is a significant
  physical phenomenon with measurable effects on the behaviour of
  $E_r$}.

{A similar discussion was made with a simplified neoclassical
  formulation in reference~\cite{itoh2007itb}, where the zonal-flow
  {neoclassical} damping was predicted to be smaller in electron
  root regimes, which was speculated to explain the enhanced
  confinement in core-electron root plasmas of some stellarators. A
  full understanding of this phenomenon would require an appropriate
  calculation of $\check{j_r}$ and also to include non-local terms in
  the drift-kinetic equation in order to study the damping of zonal
  flows. Some calculations in this direction are presented in
  Section~\ref{SEC_GK}. In the present section, within the
  neoclassical framework, we predict a peaking of the low-frequency
  $E_r$ fluctuations in the electron root of stellarators, at the
  critical density,} and we present direct experimental evidence. We
also note that the involved frequencies are the ones expected to
display higher {LRCs}, since they are
roughly below the electron transit frequency for these
plasmas. Therefore the vanishing of the neoclassical viscosity
provides a fundamental explanation of why {LRCs, and hence zonal
  flows}, are preferentially observed close before the low-density
transition of TJ-II.

Interestingly, LRCs are also observed close before the L-H transition
of TJ-II. According to figure \ref{FIG_AMB}, if one further increases
the density beyond the regime of the experiments in this work, the
slope of the ambipolar equation decreases and the low-frequency
fluctuations in $E_r$ are expected to increase again. Nevertheless,
for {those} plasmas the ambipolar equation is always similar to the one
for $n\!>\!n_{cr}$ in figure \ref{FIG_AMB}, according to our
calculations: no change of root and therefore no vanishing of $\nu_p$
is predicted.

When the density is {reduced}, the radial electric field goes
back to positive at a different critical {density.
No} vanishing of the viscosity is predicted in the backward transition
from ion to electron root, and no peaking of {low frequency $E_r$
  oscillations}. Nevertheless, these results should be taken with
caution due to some uncertainties in our calculation of the backward
transition that we discuss in {Section~\ref{SEC_BLDT}.}

{The low density transition has been observed for a large set of
  TJ-II magnetic configurations, with quantitative but no qualitative
  differences in the evolution of
  $E_r$~\cite{guimarais2008parametric}.  No dependence was found on
  $\iota$ (and therefore on the presence of low-order rational
  surfaces in the plasma), and larger configurations underwent the
  transition for smaller $\overline{n}_e$. This is consistent with a
  neoclassical discussion of the transition: in this low
  collisionality regime, the radial neoclassical fluxes do not depend
  on the rotational transform and depend strongly {on the volume
    (roughly speaking, the magnetic field strenght modulation in a
    surface labelled by the normalized radius $\rho$ is larger for
    larger volumes).} We have repeated the calculations for several
  configurations usually operated in TJ-II, and observed qualitative
  agreement with the {experiment. Along the same line of
    reasoning, no qualitative differences between configurations are
    expected in the behaviour of the neoclassical viscosity either. At
    the critical density, it will vanish; far from it, since large
    configurations will have larger $\Gamma_i(E_r)$ and
    $\Gamma_e(E_r)$, equation (\ref{EQ_NUP}) predicts smaller
    low-frequency $E_r$ fluctuations for them, assuming that every
    other parameter in the plasma is kept the same.}}

\section{The backward transition}\label{SEC_BLDT}

\begin{figure} 
  \begin{center}     
\includegraphics[angle=0,width=1.0\columnwidth]{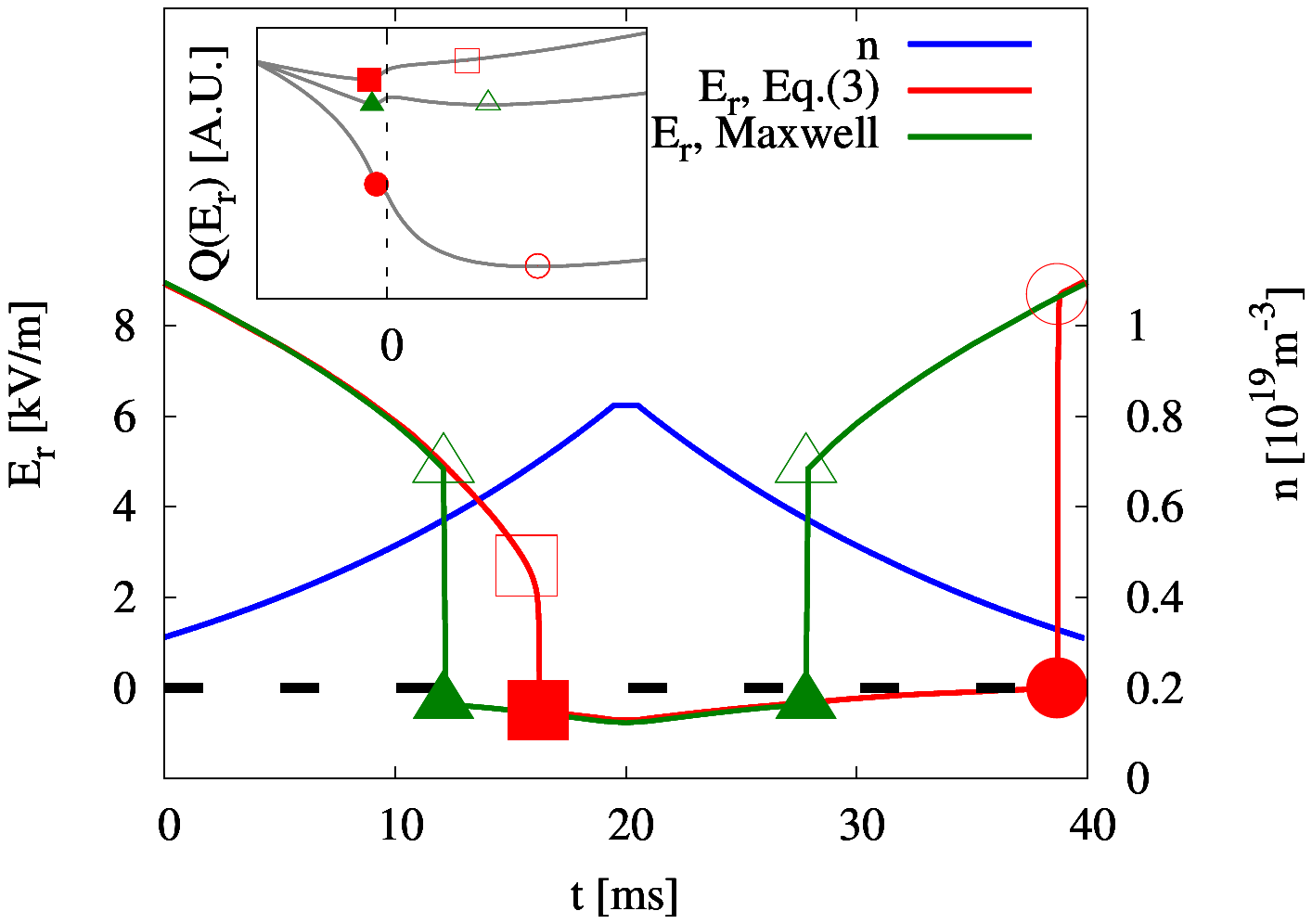} 
  \end{center} 
  \caption{Evolution of $E_r$ during the numerical density ramp for the two criteria of change of root; (inset) heat production rate for several instants of the ramp}
  \label{FIG_QER} 
\end{figure} 

\begin{figure} 
\begin{center} 
\includegraphics[angle=0]{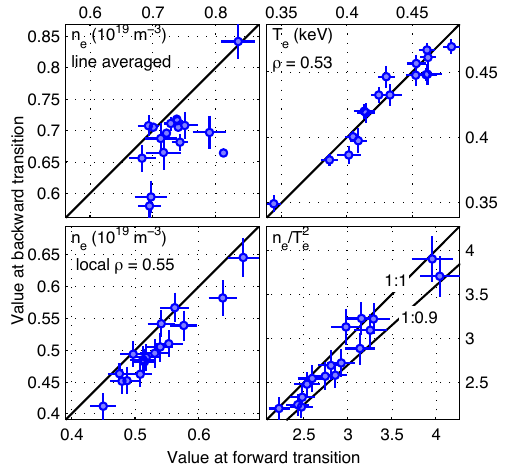} 
\end{center} 
\caption{Value at the backward transition versus value
  at the forward transition of: $\overline{n}_e$ (top-left), local
  $n_e$ (bottom-left), local $T_e$ (top-right), and local $n_e/T_e^2$
  (bottom-right).}  \label{FIG_HISTER}
\end{figure}

{Figure} \ref{FIG_EVOL} describes the evolution of $E_r$ and the
neoclassical fluxes when the density is reduced. {A very} salient
feature is that a large hysteresis in $E_r$ is predicted which is not
seen in figure \ref{FIG_TRACES}. It should be noted that this is, at
least partly, an artificial consequence of our choice of numerical
scheme. This requires a brief discussion.

In the previous section, we stated that we consider that the
transition takes place when the root in which the system is ceases to
exist. Nevertheless, from a thermodynamic point of view, one should
note that in the real system there are fluctuations and, if these are
large enough, they will make the system ``jump'' from one stable root
to the other before the first one disappears. A heuristic
thermodynamical criterion is usually employed in these kind of
calculations (see {reference}~\cite{turkin2011predictive} and references therein):
it is considered that the value of $E_r$ should be the one that
minimizes the heat production rate {defined as}
\begin{equation}
Q(E_r)\equiv \int^{E_r}[\Gamma_e-\Gamma_i](E_r')\mathrm{d}E_r'\,.
\end{equation}
It is easy to see that the minima of $Q(E_r)$ correspond to the stable
solutions of the ambipolar equation. When the latter has only one
root, both criteria yield the same result. When there are two stable
solutions, $Q(E_r)$ has two minima. According to Eq~\ref{EQ_ER}, the
transition takes place when the minimum at which the system is
disappears (open squares and closed circles in figure \ref{FIG_QER});
according to the thermodynamical argument (called {`Maxwell
construction'}) it takes place when this minimum ceases to be the
absolute minimum of $Q$ (triangles in figure \ref{FIG_QER}).

{Figure} \ref{FIG_QER} shows that our procedure of Section~\ref{SEC_NCLDT}
slightly overestimates $n_{cr}$ for the forward transition as compared
with the Maxwell construction (but it is still close to the
experimental result). Note that an accurate estimate of $n_{cr}$ with
the Maxwell criterium requires a very precise calculation of the
neoclassical fluxes close to $E_r\!=\!0$, which is not possible with
DKES{, and demands much more computing time}. On the other hand, our
procedure largely underestimates $n_{cr}$ at the backward transition,
and a large (and rather unrealistic) hysteresis is predicted by our
procedure and not by the Maxwell constraint.

Since the latter is a heuristic criterium, it is meaningful to look
more carefully at the experimental data behind figure \ref{FIG_TRACES}
and a set of similar discharges and see if some of this hysteresis is
indeed present. In order words, to check experimentally whether one of
these criteria describes perfectly the experiment or if the behaviour
of the real system lies somewhere in between the two predictions. In
figure \ref{FIG_HISTER} we show, for all the discharges of this work
with measurements at $\rho\!=\!0.55$, the value of the relevant
quantities at the backward transition versus their value at the
forward transition. The transition time is estimated from the time-lag
in the HIBP signals, as in figure \ref{FIG_TRACES}. The backward
transition systematically takes place at lower
line-densities. Nevertheless, since change of root is a radially local
feature, we focus the search of hysteresis in the local densities,
temperature and electron collisionality (which, due to the uncertainty
in the $T_i$ measurement, we take simply proportional to
$n_e/T_e^2$). The backward transition takes place at local density
around 5\% smaller and roughly at the same local $T_e$ (and hence at a
collisionality 5\% smaller). This hysteresis is of the same sign as
our calculation in figure \ref{FIG_EVOL} although much smaller.

It should be noted that, in our calculations, the plasma profiles
($n$, $T_e$ and $T_i$) for a given $\overline{n}_e$ are exactly the
same during the density ramp-up and ramp-down. This is not guaranteed
in the experiment, where we cannot measure all the thermodynamical
forces $n'/n$, $T_s'/T_s$ and $q_sE_r/T_s$ at the required times. A
careful study was made in
Refs.~\cite{happel2008twostep,milligen2011transitions} for the
density, and it was concluded that $n'/n$ was very similar at the two
crossings of $n_{cr}$. Since we are in a {regime where $E_r$ is
  mainly determined by the strong dependence of $\Gamma_e$ on the
  collisionality}, it is reasonable to assume that {the 
  differences} in the thermodynamical forces will be small enough not
  to play a relevant role. This working hypothesis is supported by
  analyses equivalent to that of figure \ref{FIG_HISTER} made (with less
  statistics) for more external radial positions in which similar
  results are observed.

Finally, it is worth noting that, according to the Maxwell
construction, the forward transition takes place slightly before the
electron root disappears (open triangle in figure \ref{FIG_QER}), and
therefore the neoclassical viscosity will reduce but not vanish (note
that even if it vanished, there would be second-order contributions to
equation (\ref{EQ_VISCO}) that would keep the fluctuations
finite). Conversely, the backwards transition will lead the plasma to
this same regime of reduced viscosity (open triangle instead of open
circle), which would explain the second peak in Fig~\ref{FIG_SPECTRA}
bottom. The fact that the second peak has usually smaller amplitude is
another experimental indication of the existence of some hysteresis.

\section{Gyrokinetic simulations of {zonal flow relaxation during} the transition}\label{SEC_GK}

{%
  In the previous sections we have shown that local neoclassical
  calculations based on the ambipolar constraint of the neoclassical
 {radial} current provide a fair understanding of the evolution of the mean
  radial electric field during {transitions associated to a change of root} in TJ-II. It was
  further shown that the viscosity $\nu_p$, defined by the linear
  dependence of those currents on $E_r$ deviations from the ambipolar
  value $E_r^0$, provides a simple explanation of the emergence of low
  frequency zonal flows near the electron-to-ion root transition and
  their reduction in fully developed ion-root plasmas.

  Zonal flow relaxation in stellarators, in the sense of the
  Rosenbluth and Hinton test~\cite{rh1998}, has an oscillatory
  character~\cite{mishchenko2008}. The calculation of the frequency of
  the oscillation, the short-time damping rate, and the residual level
  require non-local terms in the kinetic equation that are not
  included in standard neoclassical codes like DKES (see the
  discussion in the introduction of reference \cite{Satake2007}). In
  this section, and in order to address those aspects of the
  collisionless relaxation in TJ-II, we perform linear gyrokinetic
  simulations with the global, $\delta f$, particle-in-cell code EUTERPE.}

The simulations are carried out in the standard (100\_44\_64) magnetic
configuration for a density ramp similar to that of
Section~\ref{SEC_NCLDT}, based on shot \#18469. A number of $n$,
$T_e$, $T_i$ and (neoclassically calculated) $E_r$ profiles are
extracted from the ramp and introduced as an input to gyrokinetic
simulations. {Each of these simulations follows the evolution of an
  initial perturbation of the zonal flow electrostatic potential,
  $\delta\phi(\rho)$, that relaxes under linear, collisionless
  dynamics. The simulation incorporates the ambipolar radial electric
  field as a background field, so $\delta\phi'$ should be interpreted
  as a deviation from the neoclassical equilibrium value. In practice,
  the initial perturbation $\delta\phi(\rho,t=0)$ is introduced with
  an homogeneous distribution of makers whose weights vary radially as
  $w(\rho) \propto \cos(\pi\rho^2)$. This initialization of the
  markers weights gives an initial zonal perturbation to the
  distribution function and the electrostatic potential (see
  \cite{sanchez2013ishw} for more details).}

\begin{figure}
\centering
\includegraphics[width=7.7cm]{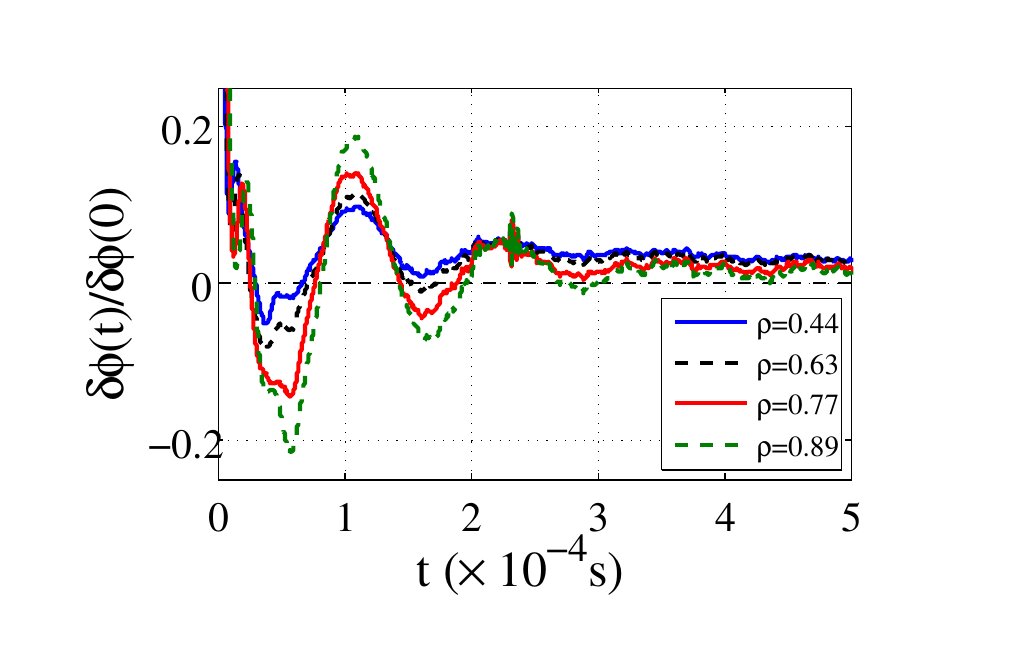}
\caption{Normalized zonal potential ($\delta\phi$ component) versus
  time at four radial positions ($\rho\!=\!$0.44, 0.63, 0.77, 0.89)}
\label{PhiOscils}
\end{figure}

{The typical relaxation of this initial perturbation, shown in}
figure~\ref{PhiOscils} {for $n\!\approx\!n_{cr}$}, shows an oscillating pattern with several
characteristic frequencies: a Low Frequency Oscillation (LFO), see
e.g.~\cite{mishchenko2008}, around $10\,$kHz and a Geodesic Acoustic
Mode (GAM)-like oscillation around $40\,$kHz.
The GAM-like oscillation {is quickly damped, while} the LFO lives for longer times.
A fit of the time traces to a {model,} 
\begin{equation}
\frac {\delta\phi(t)} { \delta\phi(0)} = A\,{\cos(2\pi f t + \Delta)} e^{-\gamma t} + R + f / (1 + k t^g)\,,
\label{FitModel}
\end{equation}
allows us to extract the frequency ($f$), amplitude of oscillations
($A$) and also the damping rate ($\gamma$) and residual level ($R$). A
simple Fourier spectrum of the time traces gives also information of
the frequency of oscillation and its amplitude {but provides no
  information about the damping rate}. The initial part
($t\!<\!25\,\mu$s) is skipped in the fit to {avoid the short-lived
  GAM oscillation}.

\begin{figure}
\centering
  \includegraphics[width=7.7cm]{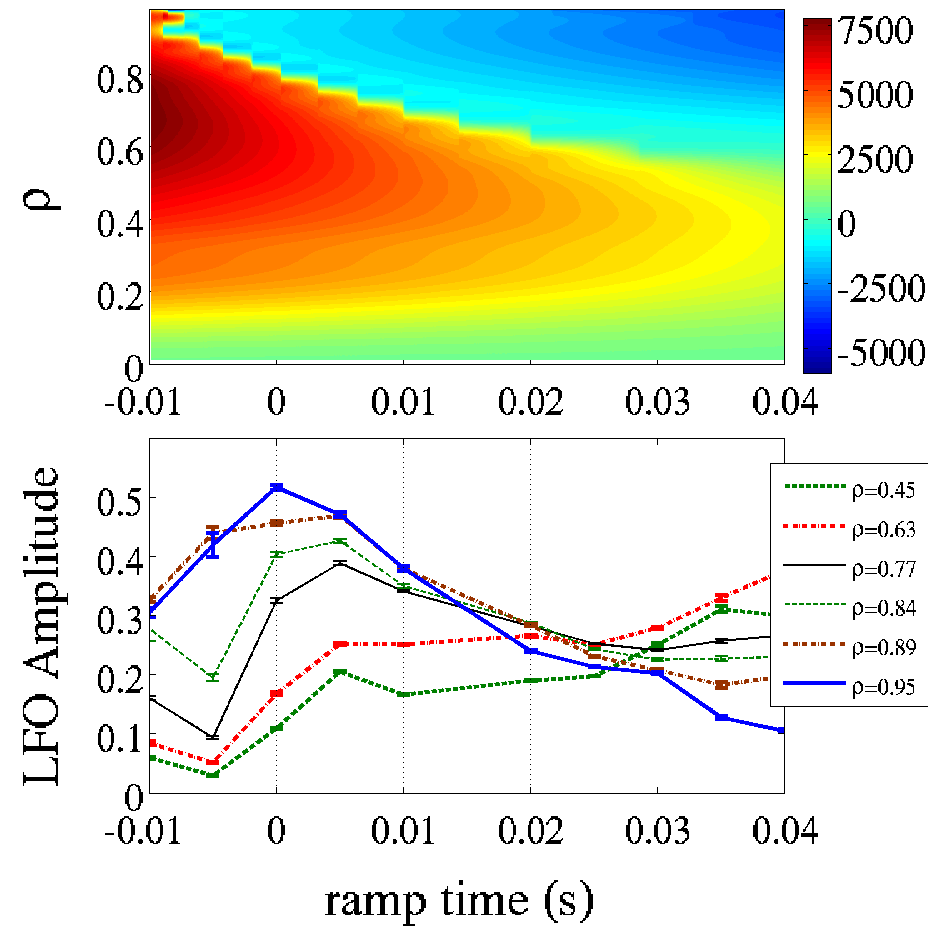}
  \caption{(top) Evolution of $E_r$ (color scale in V/m) during the
    numerical density ramp; (bottom) amplitude of the LFO obtained by
    fitting to Eq.~\ref{FitModel} for several times in the ramp up and
    several radial positions.}
\label{AmpFit}
\end{figure}

{The obtained amplitudes are shown in figure \ref{AmpFit} for
  several radial positions and several times during the ramp-up,
  together with the evolution of the neoclassical $E_r$.} The error
bars correspond to the 95\% of confidence level in the fitting.
For the outermost positions ($\rho\!\ge\!0.75$) the amplitude of the
oscillations increases as time and density increases, reaches a
maximum, and then decreases. The maximum is reached for times around
$t=0$, the time at which the neoclassical electric field is close to
zero at those radial locations. {Finite orbit widths effects of
  the ambient $E_r$ are expected to enhance the damping of the LFO in
  helical systems~\cite{mishchenko2012} in consistency with previous
  simulations~\cite{sanchez2012gkt} for the TJ-II geometry.}  {
  The maximum amplitude time delay observed for the different radial
  locations in figure \ref{AmpFit} is due to the inward propagation of
  the region of small radial electric field associated to
  electron-to-ion root jump described in previous sections. Note that
  for the simulated density ramp the radial electric field does not
  change sign in the innermost positions which consequently do not
  show a maximum in the amplitude of the {LFO.}}

  {{It} should be noted that the LFO amplitude
  increase close to the root jump observed in figure \ref{AmpFit} and
  the neoclassical viscosity reduction discussed in previous sections
  are distinct. The latter was interpreted as a reduction of the
  linear dependence of neoclassical radial currents on the radial
  electric field, whereas the gyrokinetic simulations show the effects
  of a radially varying electric potential on the finite-width ion
  orbits.  }

{ To conclude this section it is worth mentioning that the
  characteristic frequencies of the LFO ($\sim 10$ kHz) are in the
  frequency range of the electric potential fluctuations observed
  experimentally to be correlated at long
  distances~\cite{pedrosa2008prl} and interpreted as zonal-flow
  structures~\cite{alonso2012zf}. The experimental characterization of
  these structures cast a radial scale of the order of}
$k\,r_L\!=\!0.1$, where $r_L$ is the ion Larmor radius, while the
radial scale of the initial perturbation used in the simulation is
larger, $k\,r_L\!=\!0.03$ (at $\rho\!=\!0.85$, close to the radial
location of the probe measurements). Work is ongoing to study the
relaxation of perturbations with a smaller radial scale (larger $k$),
in order to compare them to the zonal flow evolution measured in
reference \cite{alonso2012zf}.

\section{Conclusions}\label{SEC_CONCL}

The main result of this work is contained in Figures~\ref{FIG_TRACES},
\ref{FIG_EVOL} and \ref{AmpFit}.  {By solving the drift kinetic
  equation, we have been able to reproduce from first principles the
  evolution of the radial electric field} during the low density
transition associated with the formation of the shear layer in
TJ-II. We {have shown how in the electron root regime of
  stellarators, close to the transition to ion root, zonal flows are
  neoclassically undamped}. {During the change of root, the mean
  radial electric field goes through zero, a situation in which our
  gyrokinetic simulations predict larger zonal flow oscillations, in
  agreement with the experiment.}

\section{Acknowledgments}

The authors thank the {TJ-II team} for their support. {They
  are also grateful to R. Kleiber and R. Hatzky for developing the
  EUTERPE code and for their continuous assistance}. This research was
{funded} in part by grant ENE2012-30832, Ministerio de
{Econom\'{\i}a y Competitividad, Spain}.

\section{Bibliography}

\bibliographystyle{plain}
\bibliography{/Users/velasco/Work/PAPERS/bibliography}


%

\end{document}